\documentclass[conference]{IEEEtran}
\IEEEoverridecommandlockouts
\usepackage{cite}
\usepackage{textcomp}
\usepackage{xcolor}
\def\BibTeX{{\rm B\kern-.05em{\sc i\kern-.025em b}\kern-.08em
    T\kern-.1667em\lower.7ex\hbox{E}\kern-.125emX}}

\usepackage[utf8]{inputenc}
\usepackage[T1]{fontenc}
\usepackage{amsmath,amssymb,amsbsy}
\usepackage{mathtools}
\usepackage{xparse}

\usepackage{graphicx}
\usepackage{dcolumn}
\usepackage{bm}
\usepackage{lipsum}
\usepackage{amsmath}
\usepackage{amssymb}
\usepackage{float}
\usepackage{setspace}

\usepackage[normalem]{ulem}
\usepackage{color}
\usepackage{caption}
\usepackage{comment}
\usepackage{array,multirow,booktabs}
\usepackage[makeroom]{cancel}
\usepackage{braket}
\usepackage{amsthm}
\usepackage{tikz}
\usepackage{mathtools}
\usepackage{comment}
\usepackage{standalone}

\usepackage{qcircuit}
\usepackage{kbordermatrix}

\renewcommand{\kbldelim}{(}
\renewcommand{\kbrdelim}{)}
\newcommand{\norm}[1]{\left\lVert#1\right\rVert}
\usepackage{enumitem}

\theoremstyle{definition}

\usepackage[colorlinks=true]{hyperref}

\title{The Quantum Alternating Operator Ansatz on Maximum $k$-Vertex Cover%
\thanks{Research presented in this article was supported by the Center for Nonlinear Studies CNLS and the Laboratory Directed Research and Development program of Los Alamos National Laboratory under project numbers 20200671DI/20190495DR.
\hfill LA-UR-19-31473
}}
\author{
	\IEEEauthorblockN{Jeremy Cook}
	\IEEEauthorblockA{\textit{Center for Nonlinear Studies CNLS} \\
	\textit{Los Alamos National Laboratory}\\
	Los Alamos, NM 87544, USA \\
	jeremycook@utexas.edu}
\and
    	\IEEEauthorblockN{Stephan Eidenbenz}
	\IEEEauthorblockA{\textit{CCS-3 Information Sciences} \\
	\textit{Los Alamos National Laboratory}\\
	Los Alamos, NM 87544, USA \\
	eidenben@lanl.gov}

\and
	\IEEEauthorblockN{Andreas Bärtschi}
	\IEEEauthorblockA{\textit{CCS-3 Information Sciences and CNLS} \\
	\textit{Los Alamos National Laboratory}\\
	Los Alamos, NM 87544, USA \\
	baertschi@lanl.gov}
}

\begin{document}
\maketitle

\begin{abstract}
The Quantum Alternating Operator Ansatz is a generalization of the Quantum Approximate Optimization Algorithm (QAOA) designed for finding approximate solutions to combinatorial optimization problems with hard constraints. In this paper, we study Maximum $k$-Vertex Cover under this ansatz due to its modest complexity, while still being more complex than the well studied problems of Max-Cut and Max E3-LIN2. 

Our approach includes (i) a performance comparison between easy-to-prepare classical states and Dicke states as starting states, (ii) a performance comparison between two $XY$-Hamiltonian mixing operators: the ring mixer and the complete graph mixer, (iii) an analysis of the distribution of solutions via Monte Carlo sampling, and (iv) the exploration of efficient angle selection strategies. 

Our results are: (i) Dicke states improve performance compared to easy-to-prepare classical states, (ii) an upper bound on the simulation of the complete graph mixer, (iii) the complete graph mixer improves performance relative to the ring mixer, (iv) numerical results indicating the standard deviation of the distribution of solutions decreases exponentially in $p$ (the number of rounds in the algorithm), requiring an exponential number of random samples to find a better solution in the next round, and (v) a correlation of angle parameters which exhibit high quality solutions that behave similarly to a discretized version of the Quantum Adiabatic Algorithm.
\end{abstract}

\maketitle

\section{Introduction}
\par
As we approach the era of quantum advantage, it is equally important to improve the quality and robustness of quantum computers as it is to improve the efficiency and noise resilience of algorithms to be run on them. Generally regarded as an ideal candidate algorithm for near term quantum computers, the Quantum Approximation Optimization Algorithm (QAOA) is a heuristic algorithm for finding approximate solutions to combinatorial optimization problems with provable approximation ratios for the problems Max-Cut~\cite{farhi2014quantum} and bounded occurrence Max E3-LIN2~\cite{farhi2014-e3lin2} using a constant circuit depth. Although the quality guarantee of the solutions are high, the QAOA with one round cannot outperform the best classical algorithm for these problems~\cite{barak2015beating}. Further, the class of problems the QAOA can solve is restricted to problems where all bitstring inputs are considered valid solutions.
\par
The Quantum Alternating Operator Ansatz (QAOA), described in a seminal paper~\cite{hadfield2019quantum}, is an extension of the algorithm that allows for optimization of combinatorial problems with hard constraints, such as problems where not all bitstrings are valid solutions. By considering more general classes of operators, the ansatz can restrict the state of the system to remain inside the subspace of valid solutions for the entire algorithm. This restriction is useful considering the dimension of the subspace of valid solutions grows exponentially small when compared to the dimension of the whole Hilbert space~\cite{nasa}. Currently, not much is known about the performance of this ansatz on optimization problems with constraints.

In this paper we aim to characterize and improve the performance of the QAOA under this ansatz for the problem of \textsc{Max $k$-Vertex Cover}, 
which has the hard constraint that all solutions must be of Hamming weight $k$: 
Given a graph $G$ and an integer $k$, find the maximum number of edges in $G$ which can be covered by a subset of $k$ vertices.
Classically, the problem admits a polynomial-time $0.92$-approximation~\cite{manurangsi2019} and is known to be NP-hard to approximate within $(1-\delta)$ for a small $\delta$~\cite{petrank1994hardness}, and UniqueGames-hard~\cite{ugc} to approximate within $0.944$~\cite{manurangsi2019}.

The performance of the QAOA on Max-$k$ Vertex Cover is dependent on many parameters: the number of vertices in the graph, the graph structure, the integer $k$, the initial state, the phase separator, the mixer, and the angle selection strategy. In Table~\ref{choices} we list the choices for each of these parameters considered in this paper. In Section~\ref{sec:qaoa} we give a brief overview of the QAOA and the Quantum Approximate Operator Ansatz. In Section~\ref{sec:formulation} we provide the problem formulation of Max-$k$ Vertex Cover in the QAOA language. In Section~\ref{sec:simulation} we show that simulation of this problem only requires keeping track of $\Theta( 2^n/\sqrt{n})$ amplitudes instead of $2^n$ amplitudes. In Section~\ref{sec:initial} we compare the performance of the QAOA between easy-to-prepare classical states and the Dicke states. In Section~\ref{sec:mixers} we define the ring mixer and the complete graph mixer and analyze their periodicity, simulation, and performance. In Section~\ref{sec:angles} we explore the distribution of solutions, optimal angle patterns, and angle selection strategies.

\begin{table*}[t]
    \centering
    \renewcommand{\arraystretch}{1.2}
    \begin{tabular}{@{}p{0.23\linewidth}p{0.23\linewidth}p{0.23\linewidth}p{0.24\linewidth}@{}}
        \toprule
        \bf{Graph Parameters}       & \bf{Initial States}       & \bf{Mixers}           & \bf{Angle Selection Strategies}   \\
        \midrule
        $|V| = 7,8,9,10$            & Dicke States              & Complete Graph Mixer  & Monte Carlo                       \\
        Pr[edge] $= 0.5$            & Random $k$-states         & Ring Mixer            & Basin Hopping                     \\
        $k = \lfloor V / 2 \rfloor$ &                           &                       & Interpolation                     \\
        \bottomrule
    \end{tabular}
    \caption{Study Parameters: We study QAOA on random graphs for different initial states and mixers, and test a variety of angle selection strategies.
    All graphs are randomly generated with 7-10 vertices and an edge connectivity probability of 0.5.}
    \label{choices}
\end{table*}

\subsection*{Related work}
While not many combinatorial problems have been studied in the QAOA setting, a few results exist. In particular, a similar analysis for Maximum $k$-colorable Subgraph problem exist~\cite{nasa} as we perform for Maximum $k$-Vertex Cover. MaximumCut QAOA has been studied extensively~\cite{farhi2014quantum,nasa}. Other problems studied include independent set~\cite{saleem2020max}, network flow~\cite{zhang2020qed}, protein folding~\cite{fingerhuth2018quantum}, and  portfolio optimization~\cite{hodson2019portfolio}. Ring mixers and complete graph mixers, which we use in our work, were introduced in~\cite{hadfield2019quantum}, and analyzed in~\cite{nasa}. Comparing $W-$states as starting states with single feasible solution starting states for graph coloring problems has been performed in~\cite{nasa}, where $W$ states are the equivalent to our more general Dicke states for $k=1$.

\section{The QAOA}\label{sec:qaoa}

\par
Given a combinatorial optimization problem over inputs $x\in \{0,1\}^n$, let $f(x)\colon \{0,1\}^n \to \mathbb{R}$ be the objective function which evaluates the cost of solution $x$. For a maximization (minimization) problem, we wish to find an $x$ for which $f(x)$ is large (small). The QAOA is specified by
\begin{itemize}
    \item An initial state $\ket{\psi}$,
    \item A `phase separating' Hamiltonian, diagonal in the computational basis: $H_P \ket{x} = f(x) \ket{x}$,
    \item A `mixing' Hamiltonian: $H_M$,
    \item An integer $p\geq 1$, the number of rounds/levels to run the algorithm, and
    \item Two real vectors $\vec{\gamma} = (\gamma_1,...,\gamma_p)$ and $\vec{\beta} = (\beta_1,...,\beta_p)$, each of length $p$.
\end{itemize}
The algorithm consists of preparing the initial state $\ket{\psi}$, and applying $p$ rounds of the alternating simulation of the phase separating Hamiltonian for time $\gamma_i$ and the mixing Hamiltonian for time $\beta_i$:
\begin{gather}
    \ket{\vec{\gamma},\vec{\beta}} = \underbrace{e^{-i\beta_p H_M} e^{-i\gamma_p H_P}}_{\text{round }p}\cdots \underbrace{e^{-i\beta_1 H_M} e^{-i\gamma_1 H_P}}_{\text{round }1} \ket{\psi}
\end{gather}
In each round, $H_P$ is applied first, which separates the basis states of the state vector by phases $e^{-i\gamma f(x)}$. The mixing operator $H_M$ then provides parameterized interference between solutions of different cost values. After $p$ rounds, the state $\ket{\vec{\gamma},\vec{\beta}}$ is measured in the computational basis and returns a sample solution $y$ of cost value $f(y)$ with probability $|\braket{y|\vec{\gamma},\vec{\beta}}|^2$. If the initial state is chosen from the subspace of valid solutions and the mixing operator acts invariantly on this subspace, all measurements are guaranteed to be valid solutions. 

The goal of QAOA is to prepare a state $\ket{\vec{\gamma},\vec{\beta}}$ from which we can sample a solution $y$ with high cost value $f(y)$.
This is usually approached by searching for angles $\vec{\gamma}$ and $\vec{\beta}$ 
such that the expectation value $\braket{\vec{\gamma},\vec{\beta}|H_P|\vec{\gamma},\vec{\beta}}$ is large ($-H_P$ for minimization problems).

Although a high expectation value of $H_P$ does not necessarily imply the measurement distribution has a high concentration of probability on optimal solutions, 
for problems with cost values in a discrete range $0,1,\ldots,m$ (such as Max $k$-Vertex Cover) this approach can be justified in the following way:
Taking $S$ sample solutions from the state $\ket{\vec{\gamma},\vec{\beta}}$, we denote their cost values by independent and identically distributed random variables $X_1,\ldots,X_S$
with expectation $\mathbb{E}[X_i] = \mu = \braket{\vec{\gamma},\vec{\beta}|H_P|\vec{\gamma},\vec{\beta}}$ and variance $\sigma^2 \leq m^2$. Then for a high enough number of samples we have the following:
\begin{itemize}
    \item   A high probability to sample at least one solution with a high cost value relative to $\mu$:
            Using Markov's inequality for random variables $Y_i = m-X_i$ with $\mathbb{E}[Y_i]=m-\mu$, and $S = m\log m$, we get
            \begin{align*}
                \mathrm{Pr}[\max X_i > \mu-1]   
                &=  1 - ( 1 - \mathrm{Pr}[X_i > \mu-1] )^S    \\
                &=  1 - ( \mathrm{Pr}[Y_i \geq m-\mu+1] )^S     \\
                &\geq 1 - (\tfrac{m-\mu}{m-\mu+1} )^S             \\
                &\geq 1-(1-\tfrac{1}{m})^S \geq 1-\tfrac{1}{m}.
            \end{align*}
    \item   A high probability to get a sample mean that is a good estimation of $\mu$: 
            Using Chebyshev's inequality for a random variable $Y = \tfrac{1}{S} \sum X_i$ with $\mathbb{E}[Y]=\mu, \mathrm{Var}[Y]=\tfrac{\sigma^2}{S}$, and $S = m^3$, we get
            \begin{align*}
                \mathrm{Pr}[\lvert Y-\mu\rvert <1]
                &= 1-\mathrm{Pr}[\lvert Y-\mu \rvert \geq 1]    \\
                &\geq 1-\tfrac{\mathrm{Var}[Y]}{1^2} \geq 1-\tfrac{\sigma^2}{S} \geq 1-\tfrac{1}{m}.
            \end{align*}
\end{itemize}
In cases where the distribution of $H_P$ is concentrated around the expectation, the number of samples can be reduced, e.g. for \textsc{MaxCut} on bounded-degree graphs with a small number of QAOA rounds~\cite{farhi2014quantum}.
Reducing the number of samples to estimate the expectation of a Hamiltonian $H$ is also important in other areas such as quantum chemistry, where $H$ is not diagonal in the computational basis~\cite{arrasmith2020}.

We now look at the behaviour of QAOA when increasing the number of levels/rounds $p$. Let
\begin{align}
    F_p(\vec{\gamma},\vec{\beta}) &=  \braket{H_P} = \braket{\vec{\gamma},\vec{\beta}|H_P|\vec{\gamma},\vec{\beta}}, \\
    M_p &= \max_{\vec{\gamma},\vec{\beta}} F_p(\vec{\gamma},\vec{\beta}).
    \label{eqn:mp}
\end{align}
Then $M_p$ has the following two properties~\cite{farhi2014quantum}:
\begin{gather}
    M_{p+1} \geq M_p, \\
    \lim_{p\to\infty} M_p = \max_{x\in\{0,1\}^n} f(x).
\end{gather}
The first property is due to the fact that $M_{p+1}$ can be seen as a constrained optimization of $M_p$, because choosing angles $\gamma_{p+1} = 0$ and $\beta_{p+1} = 0$ at level $p+1$ gives $M_p$. The second property is due to the fact that this algorithm is at least as good as an increasingly better Trotterization of the Quantum Adiabatic Algorithm~\cite{farhi2001quantum} in the limit $p\to\infty$.

\section{Problem Formulation}\label{sec:formulation}
Given a graph $G = (V,E)$, and an integer $k \in [1, |V|]$, Max-$k$ Vertex Cover is the problem of finding a set of $k$ vertices (called the $k$-set) which has the largest number of edges incident on those vertices. In order to keep the number of parameters of our study manageable, we set $k = \lfloor |V|/2 \rfloor$ throughout this paper, and leave other values of $k$ for future work.\par
Given a graph of $n$ vertices, let each vertex be represented by a binary variable $x_u$, where $x_u = 1$ if it's in the $k$-set and $x_u = 0$ otherwise. A valid solution to the problem is an $n$ bitstring $x = (x_1,...,x_n)$ of Hamming weight $k$. The objective function $f:\{0,1\}^n\to \mathbb{N}$ counts the number of edges touched by the vertices in the $k$-set, where an edge is counted if either of the two vertices on that edge is in the $k$-set. Let $\{uv\}$ denote the edge between vertex $x_u$ and $x_v$, and let $E$ be the edge set. The objective function is
\begin{align}
\begin{split}
	f(x) &= \sum_{\{uv\}\in E} \text{OR}(x_u,x_v) \\
	&= \sum_{\{uv\}\in E} 1 - (1-x_u)(1-x_v)
\end{split}
\end{align}
Substituting $(I-Z_u)/2$ for each binary variable, the phase separating (cost function) Hamiltonian is:%
\footnote{A problem mapping for Maximum $k$-Vertex Cover in the QAOA-compendium~\cite{hadfield2019quantum} is missing the linear terms of $H_P$. 
In this case, one actually counts (up to constants) the number of edges in a cut of the graph into vertex sets of size $k$ and $n-k$, rather than 
the edges covered by the set of size $k$.}
\begin{equation}
	H_P = \frac{1}{4}\sum_{\{uv\}\in E}3I - Z_uZ_v - Z_u - Z_v.
	\label{phaseop}
\end{equation}
The period of $H_P$ is at most $2\pi$ because all the eigenvalues of $H_P$ (image of the cost function) are integers.

\section{Simulation}\label{sec:simulation}
An advantage of the Quantum Alternating Operator Ansatz over the QAOA is that the quantum state vector remains in the subspace of valid solutions during the whole circuit, so there is zero probability of measuring an invalid solution (provided there is no noise or error in the circuit). In our case, the subspace of valid solutions is the space spanned by all Hamming weight $k$ states. The phase separator and the mixing operators listed in Table~\ref{choices} all act invariantly on any Hamming weight $k$ subspace. The phase separator acts invariantly because it is diagonal. The Hamiltonians $H_M$ of both the complete graph mixer and ring mixer are $XY$-Hamiltonians, meaning they are sums of $XY$ terms:
\begin{gather}
    X_iX_j + Y_iY_j = 
    \kbordermatrix{
        & \ket{0_i 0_j} & \ket{0_i 1_j} &  \ket{1_i 0_j} & \ket{1_i 1_j} \\
        \bra{0_i 0_j} & 0 & 0 & 0 & 0 \\
        \bra{0_i 1_j} & 0 & 0 & 1 & 0 \\
        \bra{1_i 0_j} & 0 & 1 & 0 & 0 \\
        \bra{1_i 1_j} & 0 & 0 & 0 & 0 
    },
\end{gather}
where the sums range either over all neighboring pairs of qubits (with periodic boundary) in the case of the ring mixer (cf.~Equation~\eqref{eq:xy-ring-mixer}), or over all pairs of qubits in case of the complete graph mixer (cf.~Equation~\eqref{eq:xy-completegraph-mixer}).  

A 2-qubit $XY$-Hamiltonian acts as the SWAP operator in the subspace $\{\ket{0_i1_j},\ket{1_i0_j}\}$, which preserves Hamming weight, and otherwise destroys the state. A sum of these operators acting on any pair of qubits preserves Hamming weight by linearity, so both the complete graph mixer and ring mixer act invariantly on any Hamming weight $k$ subspace. 
Provided the initial state is a superposition of Hamming weight $k$ states, the entire circuit can be simulated in this subspace, so there are only $\binom{n}{k}$ amplitudes to keep track of, instead of $2^n$. 
For constant $k$ this simulation is efficient and uses $\Theta(n^k)$ amplitudes, but for $k=\lfloor n/2 \rfloor$, this simulation requires a state vector of size $\Theta(2^n/\sqrt{n})$ by bounds related to the Stirling approximation.

\section{Initial state}\label{sec:initial}

We compare the performance of two different initial states: Dicke states and random $k$-states. The Dicke state is an equal superposition of all Hamming weight $k$-states:
\begin{equation}
    \ket{D_k^n} = \frac{1}{\sqrt{\binom{n}{k}}}\sum_{|x|=k}\ket{x}.
\end{equation}
Because the complete graph mixer and ring mixer preserve Hamming weight, the Dicke states are eigenvectors of these mixers by symmetry for all $k$.
Further, for $k= \lfloor n/2 \rfloor $, the Dicke state is the ground state of both the complete graph mixer and the ring mixer. For any $k$, the Dicke state can be prepared with $O(kn)$ gates in $O(n)$ depth without ancilla qubits, even on Linear Nearest Neighbor (LNN) architectures~\cite{bartschi2019dicke}.
The construction is inductive and improves in the mentioned metrics on previous approaches that used arithmetic circuits~\cite{Mosca2001}, probabilistic preparation with a projective measurement~\cite{Childs2002} or circuits with exponential scaling for super-constant $k$~\cite{wang2009}.
An exception is a construction for $W$-states with $O(n)$ gates that has $O(n)$ depth on LNN architectures and $O(\log n)$ depth on tree-shaped hardware connectivity~\cite{Wstates-logdepth}.

\begin{figure*}[t]
    \centering
        \hspace*{0.075\linewidth}Complete Graph Mixer
        \hspace*{0.2\linewidth}Ring Mixer
        \hfill\hfill
        Mixer Comparison
        \hspace*{0.05\linewidth} \\
        \includegraphics[width=0.3\linewidth]{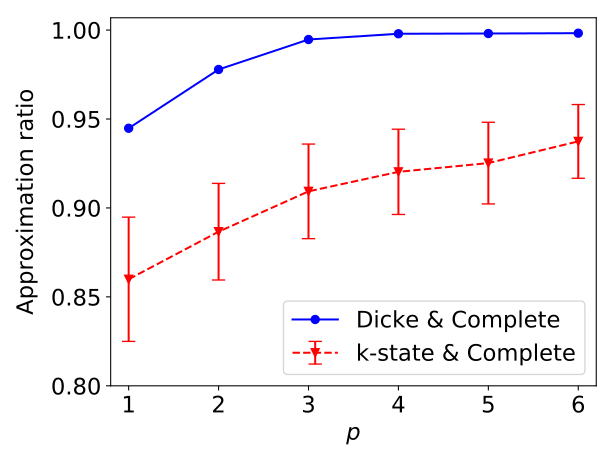}
        \hfill
        \includegraphics[width=0.3\linewidth]{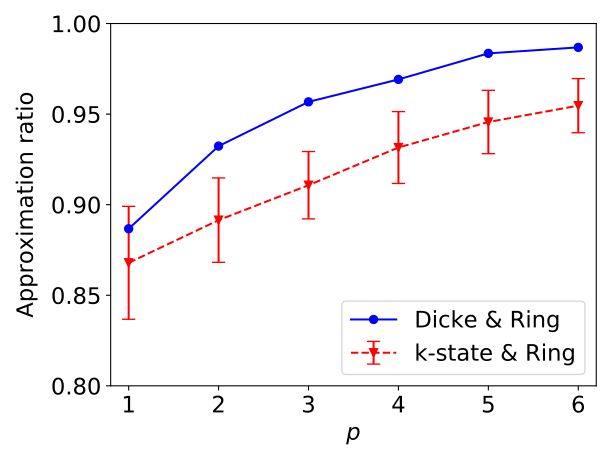}
        \hfill\hfill
        \includegraphics[width=0.3\linewidth]{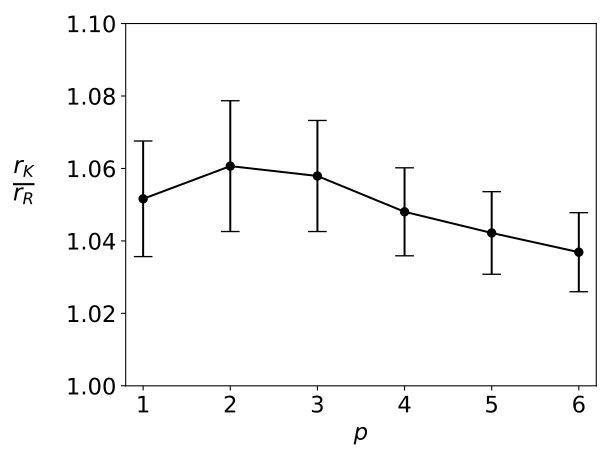}
    \caption{\textbf{(left)} The relative performance of the Dicke states compared to the average over all random $k$-states as the initial state (giving confidence intervals), using the complete graph mixer and ring mixer. Plots show result of a representative graph from a total of 100 randomly selected graphs between 7 and 10 vertices and using optimal angles, which all looked very similar. Starting with Dicke states enables QAOA to find near-optimum solutions faster.\newline
    \textbf{(right)} The relative performance of the complete graph mixer and the ring mixer when starting in the Dicke state, averaged over 100 graphs (of size 7), where $r_K$ is the approximation ratio using the complete graph mixer, and $r_R$ is the approximation ratio using the ring mixer. The average of many samples of $r_K/r_R$ is what is shown in the graph, including a confidence interval on this estimate. The complete mixer performs better, since the ratio is above one.}
    \label{fig:initial-vs-dicke}\label{fig:mixer-comparison}
\end{figure*}

A random $k$-state is a random computational basis state of Hamming weight $k$, which can be prepared in constant depth by applying the $X$ gate to $k$ random qubits. The random choice is only made once at the beginning of the algorithm, and not during every sample of $\braket{H_P}$. If we were to randomly choose $k$ qubits to flip for every sample, we would be optimizing over the mixed state
\begin{gather}
    \rho = \frac{1}{\binom{n}{k}} \sum_{|x|=k} \ket{x}\bra{x},
\end{gather}
which in the Hamming weight $k$ subspace is proportional to the identity matrix and commutes with the entire circuit. For any angles $\vec{\gamma}, \vec{\beta}$, the expectation value for this initial state would be a constant.
\par
We measure the performance of the QAOA using the approximation ratio, defined formally as 
\begin{gather}
    \frac{F_p(\vec{\gamma},\vec{\beta})}{\max_{x\in\{0,1\}^n} f(x)},
\end{gather}
for a specific choice of initial state and mixer. The relative performance between the Dicke states and the classical $k$-states are shown in Figure~\ref{fig:initial-vs-dicke}~(left) on a representative graph for both the complete graph mixer and the ring mixer. By representative we mean the trend shown in Figure~\ref{fig:initial-vs-dicke} was similar across 100 different random graphs. For the complete graph mixer, it is clear the Dicke state outperforms the classical $k$-states. The average of the classical $k$-states do not achieve a comparable approximation ratio to the Dicke states until many rounds in. For the ring mixer, the Dicke state also performs better than the classical $k$-states, but the advantage is not quite as clear. Overall, we can be confident that the Dicke states improve performance.

We remark that when starting in a classical $k$-state, the first level phase separator $e^{-i\gamma_1 H_P}$ only acts with a global phase, 
hence only the mixer $e^{-i\beta_1 H_M}$ plays a role in the first QAOA level. However, a $p$-level QAOA starting with Dicke states still compares favorably with 
a $(p+1)$-level QAOA starting in a random $k$-state, even though the latter has an additional mixer layer that can be tuned. 
This is especially noteworthy as on LNN architectures, Dicke States can be implemented in $O(n)$ depth, 
the same depth as (as we will discuss later) the ring mixer and smaller depth than the complete graph mixer. 

The confidence intervals on the $k$-state lines are for the distributions taken over all possible $k$-states of the graph. These numerical simulations were done with an outer-loop that identified the optimum angles through a fine grid with steps size $0.01 \pi$, thus requiring significant computational resources. Our simulations were limited to small graphs (between 7 and 10 vertices) due to computational simulation resource constraints; it appears plausible that we would see similar dominance of the Dicke starting states, perhaps even more pronounced, for larger graphs on an actual error-corrected quantum computer as both mixers will take -- on average -- even longer to reach the initial equal distribution that the Dicke state starts with.


\section{Mixers}\label{sec:mixers}
Two mixing operators for fixed Hamming weight problems were presented in~\cite{nasa}: the ring mixer and the complete graph mixer. For both we will discuss the relevant search space of the parameters $\vec{\gamma}$ and $\vec{\beta}$, and the complexity of implementation. We compare the performance of the two using the Dicke state as the initial state in Figure~\ref{fig:mixer-comparison}~(right).

\begin{figure*}
    \begin{tabular}{ccc}
        5 vertices & \hspace{-1.0cm} 6 vertices & \hspace{-1.0cm} 7 vertices\\
        \hspace{-.5cm} \includegraphics[scale=0.4]{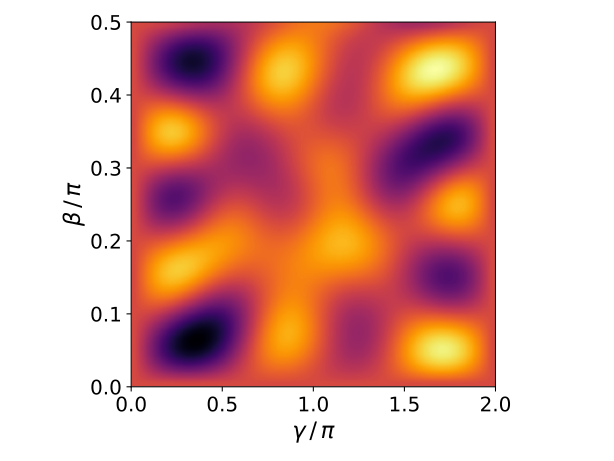}&
        \hspace{-1.0cm} \includegraphics[scale=0.4]{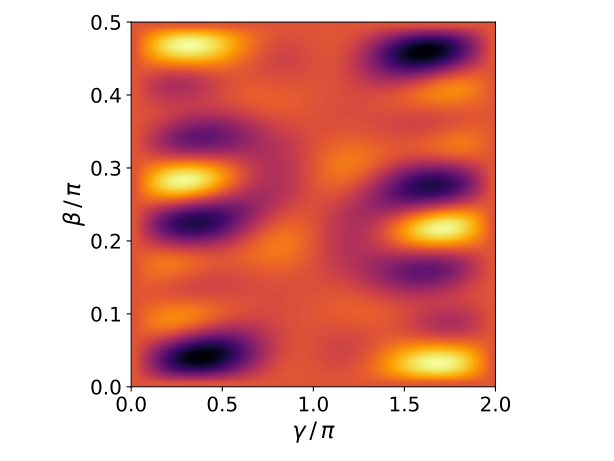} &
        \hspace{-1.0cm} \includegraphics[scale=0.4]{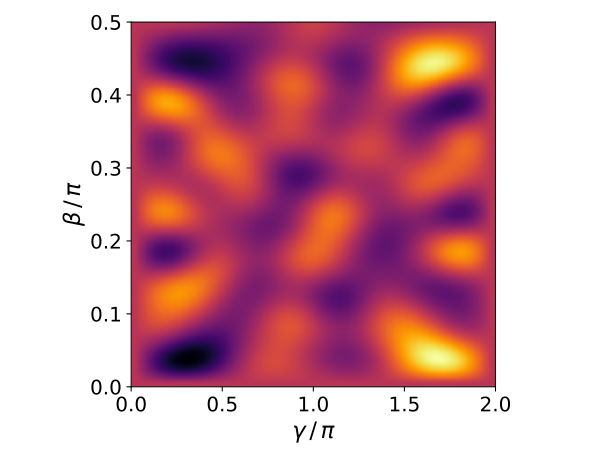} \\
        8 vertices & \hspace{-1.0cm} 9 vertices & \hspace{-1.0cm} 10 vertices\\
        \hspace{-.5cm} \includegraphics[scale=0.4]{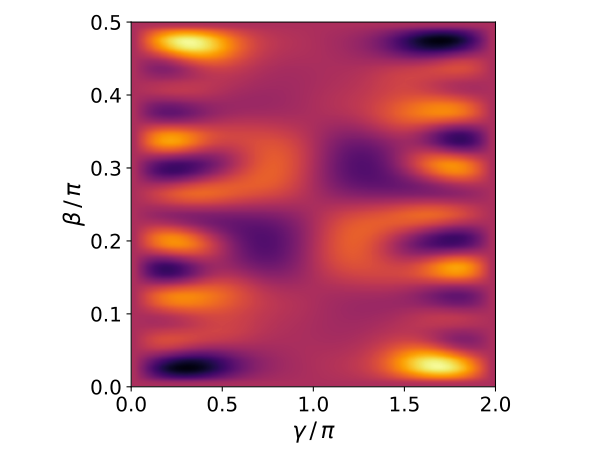} &
        \hspace{-1.0cm} \includegraphics[scale=0.4]{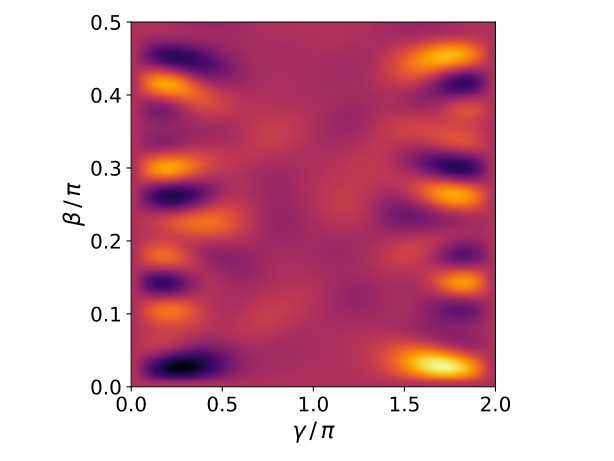} &
        \hspace{-1.0cm} \includegraphics[scale=0.4]{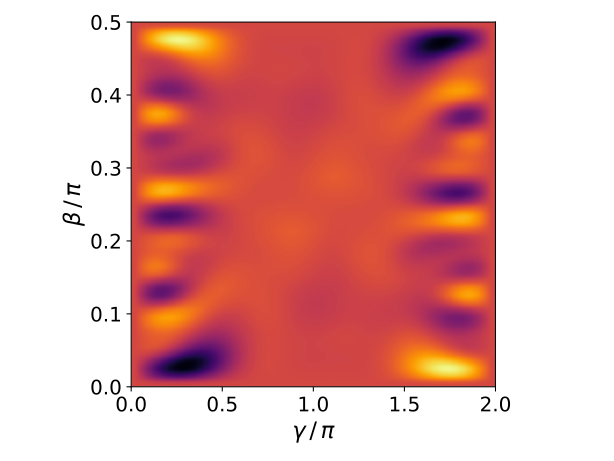}

    \end{tabular}
    \caption{Heat plot distribution representing the value of $\braket{H_P} = \braket{\vec{\gamma},\vec{\beta}|H_P|\vec{\gamma},\vec{\beta}}$ (our objective function) 
    for 6 random graphs each  of size 5 to 10 vertices, using a 1-round QAOA with the complete graph mixer and an initial Dicke state.
    Dark colors map to a high expectation value, and warm colors map to a low expectation value.}
    \label{fig:examples}
\end{figure*}

\subsection{Ring Mixer}
The $2^n \times 2^n$ ring mixer Hamiltonian is
\begin{gather}
    H_R = \sum_{i=0}^n X_i X_{i+1} + Y_i Y_{i+1}, \label{eq:xy-ring-mixer}
\end{gather}
where the indices $i+1$ are taken mod $n$.\par
The search space of the QAOA is taken over the period of the phase separating Hamiltonian for $\gamma$, and over the period of the mixing Hamiltonian for $\beta$, provided both the Hamiltonians are periodic. However, we will show the ring mixer does not have a definite period for all $n$. A Hamiltonian has a period of $x$ if $e^{-i x H} = e^{i\phi} I$, where $e^{i\phi}$ signifies an arbitrary global phase. For any set of basis vectors $B$, this can be rewritten as
\begin{gather}
    \forall \ket{v}\in B, \quad (e^{-ix H} - e^{i\phi} I)\ket{v} = 0,
\end{gather}
which is the eigenstate equation for a unitary operator. In other words, $e^{-ixH}$ has a period of $x$ if there exists an $x$ such that all the eigenvalues of $xH$ are equal modulo $2\pi$. For any $n$ the ring mixer has the eigenvectors and eigenvalues:
\begin{align}
    H_R\ket{0^n} &= 0\ket{0^n}, \\
    H_R\ket{D_1^n} &= 4\ket{D_1^n}.
\end{align}
This restricts the period $x$ to $\pi d/2$ for some $d\in \mathbb{Z}$. It can be checked analytically that for $n=4$ up to $n=10$ the ring mixer also has irrational eigenvalues, so there does not exist an $x$ for which all eigenvalues can be equal modulo $2\pi$. Then for at least $n \in [4,10]$, the ring mixer has no definite period. Without a definite period the search space cannot be defined exactly and the full range of $F_p(\vec{\gamma},\vec{\beta})$ cannot be explored. However, we retain the properties of the QAOA which are relevant. The property $M_{p+1} \geq M_p$ holds because $M_{p+1}$ is still constrained optimization of $M_p$. Further, for any constant search space $[0,C]$, in the limit $p\to\infty$, the higher order Trotterization of the Quantum Adiabatic Algorithm implies both $\gamma$ and $\beta$ are small and less than $C$, so the property $\lim_{p\to\infty}M_p = \max_{x\in\{0,1\}^n} f(x)$ also holds. That being said, taking $C$ too small can result in terrible performance, and should be taken greater than a certain threshold which contains many high quality solutions, as will be shown in Section~\ref{sec:angles}.
\par
As shown in~\cite{nasa}, the ring mixer is analogous to the one dimensional spin-1/2 chain model, which can be exactly diagonalized using the Jordan-Wigner and the Fast Fermion Fourier Transform (FFFT). The FFFT and thus the ring mixer can be implemented exactly on a quantum computer in $O(\log n)$ depth assuming all-to-all connectivity of qubits, and in 
$O(n)$ depth assuming only a LNN connectivity.

\subsection{Complete Graph Mixer}
The $2^n \times 2^n$ complete graph mixer Hamiltonian is
\begin{gather}
    H_K = \sum_{(i,j)\in E(K_n)} X_i X_j + Y_i Y_j, \label{eq:xy-completegraph-mixer}
\end{gather}
where $E(K_n)$ is the edge set of the complete graph $K_n$ of $n$ vertices. Let $\ket{y}$ be a bitstring of Hamming weight $k$, and let $S$ denote the set of all Hamming weight $k$ bitstrings which are a Hamming distance two away from $\ket{y}$. The action of the complete graph mixer on $\ket{y}$ is 
\begin{gather}
    H_K\ket{y} = 2\sum_{x\in S}\ket{x}.
\end{gather}

From the fact that $H_K$ preserves Hamming weight, it must act invariantly on the $n+1$ Hamming weight $k$ subspaces $W_k = \text{span}\{\ket{x}:|x| = k\}$. Grouping the basis vectors by Hamming weight, the matrix representation of $H_K$ is this basis is block diagonal. Each of these $\binom{n}{k} \times \binom{n}{k}$ block matrices on the diagonal represent the adjacency matrix of the Johnson graph $J(n,k)$. 
\par
The vertices of the Johnson graph $J(n,k)$ are the $k$-element subsets of an $n$-element set, where two vertices are connected if their intersection contains $k-1$ elements. Labeling the $n$ qubits $\{0,1,...,n-1\}$, states of Hamming weight $k$ can be represented by subsets of cardinality $k$, where the indices in the subset represent which qubits are $\ket{1}$, with the remaining qubits set to $\ket{0}$. Two states of Hamming weight $k$ are a Hamming distance of two apart if they contain $k-1$ elements in common, which is exactly the case of the Johnson graph. 
\par
The eigenvalues of $H_K$ are exactly the eigenvalues of the blocks on the diagonal.
The eigenvalues of the Johnson graph are $(k-j)(n-k-j)-j$ with multiplicities $\binom{n}{j} - \binom{n}{j-1}$, for $j = 1,2,...,\min\{k, n-k\}$~\cite{johnson-eigenvalues}. All of these eigenvalues are integers, so all the eigenvalues of $H_K$ are even integers. 
Hence the period of the complete graph mixer is at most $\pi$.
\par

The period of $H_P$ is $2\pi$ and the period of $H_K$ is $\pi$, so
\begin{gather*}
    F_p(2\pi-\gamma,\pi-\beta) = F_p(-\gamma,-\beta) = F_p(\gamma,\beta),
\end{gather*}
where the last equality comes from the fact that both $H_P$ and $H_K$ are real valued and satisfy time reversal symmetry. We can therefore take our search space to be $[0,2\pi)^p\times [0,\pi/2)^p$.\par

It was shown in~\cite{nasa} that the complete graph mixer in the Hamming weight $k=1$ (and symmetrically for $k=n-1$) subspace can be exactly implemented using a clever partitioning and Trotterization of the $XY$ terms. However, this Trotterization approximation is not exact in the larger Hamming weight subspaces.


Instead of implementing the mixer exactly, we look at Hamiltonian simulation techniques. Note that $H_K$ is efficiently row computable and sparse, so it can be simulated efficiently in the black-box model. In this model as defined in~\cite{berry2009black}, the matrix elements of the $N \times N$ Hamiltonian $H$ are retrieved from an oracle:
\begin{gather}
    O_H\ket{j,k,z} = \ket{j,k,z\oplus H_{jk}},
\end{gather}
for $j,k\in [N]$. Suppose our Hamiltonian is $D$ sparse, meaning there are at most $D$ non-zero elements in any row or column. Define a second oracle to calculate the row index of the non-zero elements in column j of $H$:
\begin{gather}
    O_F\ket{j,k} = \ket{j,f(j,k)},
\end{gather}
for $j\in [N]$ and $k\in [D]$, where the function $f(j,k)$ returns the row index of the $k$th non-zero element in column $j$. In such a model, the complexity of simulating the Hamiltonian is measured by the number of queries made to $O_H$ and $O_F$. Let $\norm{H}$ be the spectral norm of $H$ and $\norm{H}_{\text{max}} = \max_{j,k} \lvert H_{jk} \rvert$. Then the evolution of $H$ with sparsity $D$ can be simulated for time $t$ with error at most $\delta \in [0,1)$ using
\begin{gather}
    O\left(\frac{\norm{H}t}{\sqrt{\delta}} + \norm{H}_\text{max} Dt\right),
\end{gather}
queries to the black-boxes $O_F$ and $O_H$~\cite{berry2009black}.\par
For the complete graph mixer with $k=\lfloor n/2 \rfloor$, $D=n^2/4$, and the largest eigenvalue is $\norm{H} = k(n-k) = O(n^2)$. Furthermore, the complete graph mixer is periodic, so $t$ is a constant, and all the matrix entries of $H_K$ are 2, so $\norm{H}_\text{max}$ is also a constant. Therefore the query complexity of the complete mixer is $O(n^2/ \sqrt{\delta})$.

\begin{figure}
    \centering
    \begin{tabular}{c}
        Complete Graph Mixer \\
        \includegraphics[width=\linewidth]{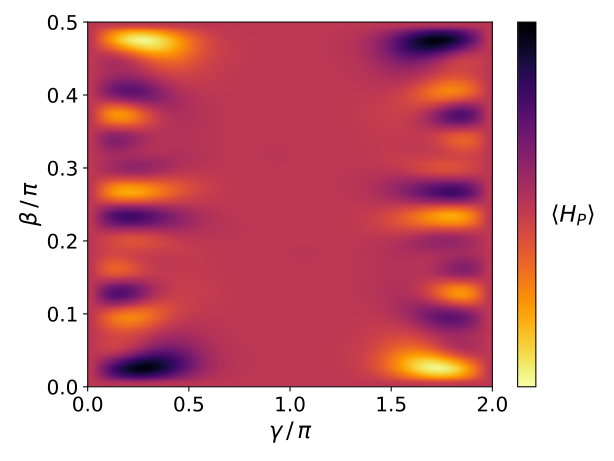} \\
        Ring Mixer \\
        \includegraphics[width=\linewidth]{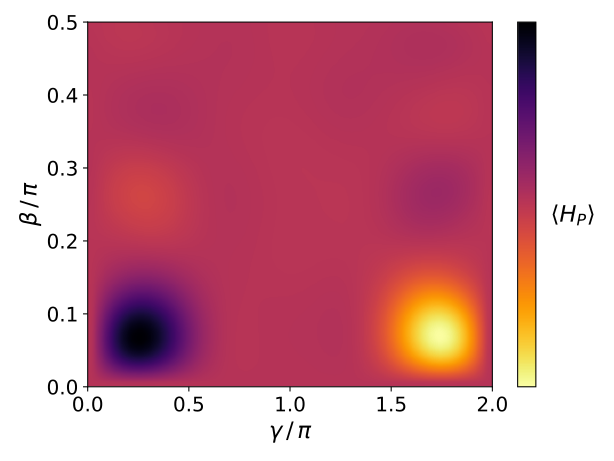} 
    \end{tabular}
    \caption{Averaged heat plot of $\braket{H_P}$ in a 1-round QAOA with initial Dicke state, using the complete graph mixer \textbf{(top)} and ring mixer \textbf{(bottom)}. Plot averaged over 100 random graphs of 10 vertices, darker colors map to a high expectation value.}
    \label{fig:avg-heat}
\end{figure}

\subsection{Mixer comparison}
To get an idea of the relative performance between the two mixers, we compare the optimal solutions ($M_p$ in (\ref{eqn:mp})) of each mixer in each round. As this comparison is highly graph dependent, we average this comparison over many random graphs. To increase the confidence of finding the optimal angles in the search space, we use smaller graphs of seven vertices so that optimization can be run quickly and repeated to verify agreement of the optimal angles. Let $r_K$ be the optimal approximation ratio of the complete graph mixer, and let $r_R$ be the optimal approximation ratio of the ring mixer. To provide a fair comparison, we restrict the optimizer to take the same number of steps when computing $r_K$ as $r_R$. The average of the ratio $r_K/r_R$ over many random graphs for multiple rounds of the algorithm is shown in Figure~\ref{fig:mixer-comparison}~(right). 
From the fact that $r_K/r_R > 1$, we can be fairly confident in saying the complete graph mixer achieves a better approximation ratio than the ring mixer, but that this advantage diminishes for larger $p$. This result is expected as the complete mixer by design distributes amplitudes across all valid solutions in each round, whereas the ring mixer takes several rounds to completely mix and indeed approaches the same performance if we go to large $p$. While the single-digit percent advantage we show in the plot may not be that impressive, we believe that the advantage of the complete mixer increases with larger graphs as the ring mixer will take longer and longer to mix adequately.

\emph{For this reason, we will mainly study properties of the complete graph mixer for the remainder of the paper.}
For a parameter landscape of a 1-round QAOA with a complete graph mixer and an initial Dicke state, see Figure~\ref{fig:examples}, for 
an averaged landscape comparison between complete graph mixer and ring mixer, see Figure~\ref{fig:avg-heat}.

Both the ring mixer and the complete graph mixer can be implemented efficiently on a quantum computer. Although the complete graph mixer achieves a better optimal approximation ratio, in practice this advantage may be overshadowed by the implementation depth of $O(n^2/\sqrt{\delta})$ compared to $O(\log n)$ of the ring mixer (assuming full connectivity).

\section{Angle Selection}\label{sec:angles}
One of the main challenges of the QAOA is efficiently finding optimal or near-optimal angles $\vec{\gamma}$ and $\vec{\beta}$. As is the case for Max-Cut and bounded occurrence Max E3-LIN2, the QAOA with $p = 1$ does not perform as well as the best known classical algorithms for these problems, making it necessary to study the behavior of the QAOA for $p>1$ if the algorithm is to compete with its classical counterparts.

There has been significant work in QAOA parameter setting over the past few years. The original proposal of using a fine grid search~\cite{farhi2014quantum} scales exponentially in the number of levels $p$. 
Later work has looked at QAOA parameter search based on gradient descent~\cite{wang2018fermionic}, basin-hopping~\cite{nasa,olson2012basin}, or interpolation of angles for a level-($p+1$) QAOA based on good angles for $p$ levels~\cite{zhou2018quantum}. 
The latter makes use of the observation that the phase separator angles $\gamma_i$ often monotonically increase while angles $\beta_i$ decrease, which 
can be seen as a digitized version of quantum annealing~\cite{mbeng2019quantum}. Exploring the connections between quantum annealing and QAOA, 
there is discussion in which cases the optimal control of the two Hamiltonians $H_P, H_M$ is bang-bang~\cite{yang2017optimizing} (such as in QAOA) or a combination of bangs and anneals~\cite{brady2020optimal}. Recently considered mixing unitaries based on Grover's diffusion operator~\cite{biamonte2020,baertschi2020} might give a similar connection to fixed-point quantum search~\cite{grover_fixed_point,yoder2014fixed}.

Before we get to the angle selection strategies used in this paper, we briefly consider the total gate complexity of the QAOA for $p$ rounds, which is
\begin{gather}
    f(i + p(s+m)), \label{eqn:complexity}
\end{gather}
where $i$ is complexity of preparing the initial state, $s$ is the complexity of implementing the phase separator, $m$ is the complexity of implementing the mixer, and $f$ is the complexity of finding optimal angles based on samples of $\braket{H_P}$. In practice there is the added complexity of estimating $\braket{H_P}$, but for simplicity we will assume this to be constant.

In the case of Max-Cut, $p$ does not grow with $n$ in the calculation of $\braket{H_P}$ due to the locality of the mixing operator and the regularity of the graph, causing many terms in $\braket{H_P}$ to commute and cancel. However, this is not true in the general case with non-local mixing operators and non-regular graphs. In  the case of Max-$k$ Vertex Cover, the operator $e^{-i\beta H_M}$ must be exponentiated explicitly in order to calculate $\braket{H_P}$ classically, which quickly becomes intractable as $n$ grows large. To be general, we will assume $p$ to grow polynomially in $n$, so that to keep the complexity of (\ref{eqn:complexity}) efficient in $n$, $f$ must be a polynomial in $p$. The problem is that for the search domain $[0,2\pi)^p \times [0,\pi/2)^p$, the volume of this space grows exponentially with $p$, so we cannot efficiently sample from a fine grid of the whole search space. A different angle selection strategy must be employed to find optimal angle solutions efficiently. 

In the following, we develop and assess three different angle selection strategies, focusing on the complete graph mixer and initial Dicke states.
To get a sense of what the distribution of expectation values $\braket{H_P}$ look like at level $p=1$ as a function of the two angles $\beta, \gamma$, in Figure~\ref{fig:examples} we have plotted example distributions over the search space for random graphs ranging between 5 and 10 vertices.

\begin{figure*}[t]
    \centering
        \includegraphics[width=0.32\linewidth]{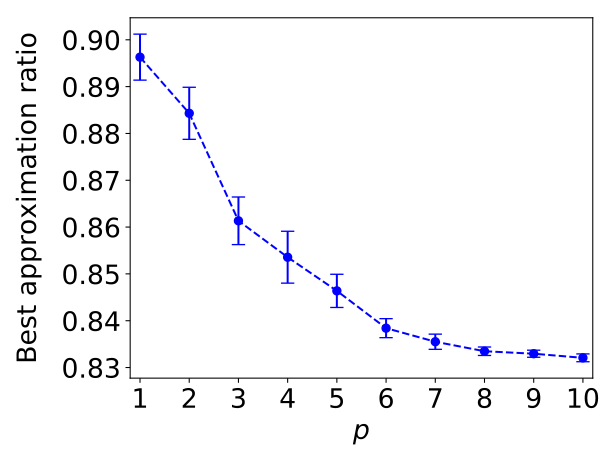}\hfill
        \includegraphics[width=0.32\linewidth]{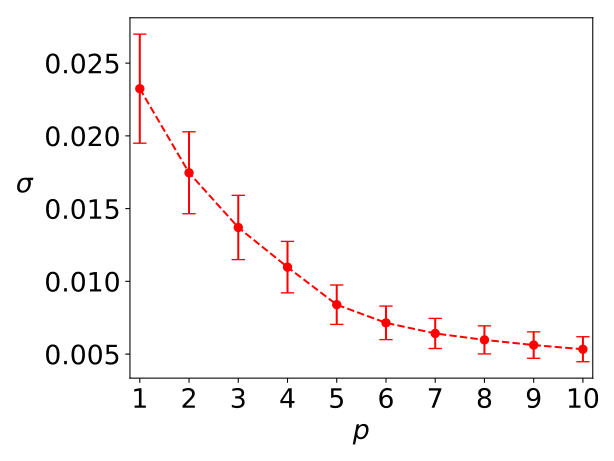}\hfill
        \includegraphics[width=0.32\linewidth]{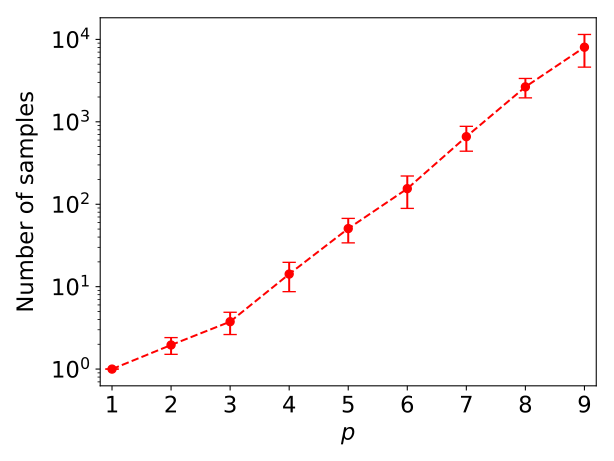}
    \caption{Monte Carlo Sampling for angle selection. 
    \textbf{(left)} The best found solution as an approximation ratio at each level.
    \textbf{(middle)} An estimate of the standard deviation of expectation values at each level. 
    For both plots a constant number of 1000 samples were taken at each level to produce an estimate, and this was repeated 1000 times to produce a confidence interval. As performance decreases with higher $p$ counts, Monte Carlo Sampling is not a promising candidate for angle selection.\newline
    \textbf{(right)} The number of Monte Carlo samples required to find a better solution at each level increases exponentially in $p$.}
    \label{fig:monte-std}\label{fig:better_sol}
\end{figure*}

\subsection{Monte Carlo Sampling}
In the distribution of $\braket{H_P}$, we are mainly interested in the upper tail of the distribution corresponding to optimal and near-optimal solutions. Instead of sampling from a compact grid over the whole parameter space, whose run time grows exponentially with $p$, our first angle selection strategy is straight-forward Monte Carlo sampling: randomly select each component of the two vectors $ \vec{\gamma} = (\gamma_1,...,\gamma_p)$ and $\vec{\beta} = (\beta_1,...,\beta_p)$. We take a constant number of such Monte Carlo samples to investigate properties of the distribution. 

\par
From random samples of $\braket{H_P}$, we collected the best found solution and an estimate of the standard deviation of the distribution. This procedure was repeated to get a confidence interval on these estimates, with the results plotted in Figure~\ref{fig:monte-std}~(left/middle). This data was collected on a representative graph of 10 vertices, where by representative we mean the observed patterns were also observed in many other trials of random graphs. (We used Dicke states as starting states and the complete mixer for these experiments.) The plot in Figure~\ref{fig:monte-std}~(middle) shows an exponentially decreasing standard deviation, and the plot in Figure~\ref{fig:monte-std}~(left) shows an exponentially decreasing quality of solution. 
These two plots imply that the average quality of solutions is not increasing at a fast enough rate for the tail end of the distribution to increase past the previous round, due to the decreasing standard deviation. For increasing $p$, higher quality solutions become harder and harder to find. To drive this point home, the number of random samples required to find a better solution at level $p+1$ than in level $p$ grows exponentially in $p$, as shown in Figure~\ref{fig:better_sol}~(right), with a logarithmic scale on the $y$-axis. Thus, given a constant budget of Monte Carlo sample runs, it would be best to limit the number of rounds to $p = 1$, which defies the true power of any QAOA approach. Naive Monte Carlo sampling does not appear to be a promising approach to an angle selection strategy.

\subsection{Basin Hopping}
The rough landscape at level one of the QAOA with many hills and valleys in Figure~\ref{fig:examples} suggests using a gradient descent optimization algorithm which randomly hops out of local minima often to find a better minimum. For this reason, we use a basin-hopping algorithm when searching for optimal angles. The basin-hopping algorithm locally optimizes a solution through gradient descent and then perturbs the solution more significantly to find a new solution. While often used and originally proposed for molecular biology applications~\cite{olson2012basin}, basin hopping has been used for other QAOA problems as well~\cite{nasa}. Our basic basin hopping approach starts with a random angle vector. We present results in the next section comparing basic basin hopping to a more advanced version of basin hopping.

\begin{figure*}
    \centering
    \begin{tabular}{ c c }
         & \\
        \includegraphics[width=200pt]{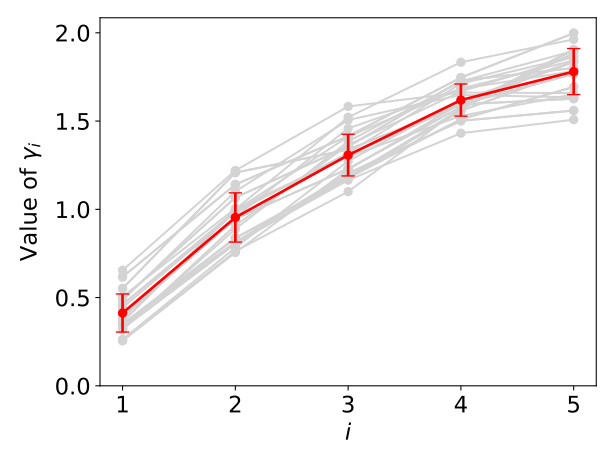} &
        \includegraphics[width=200pt]{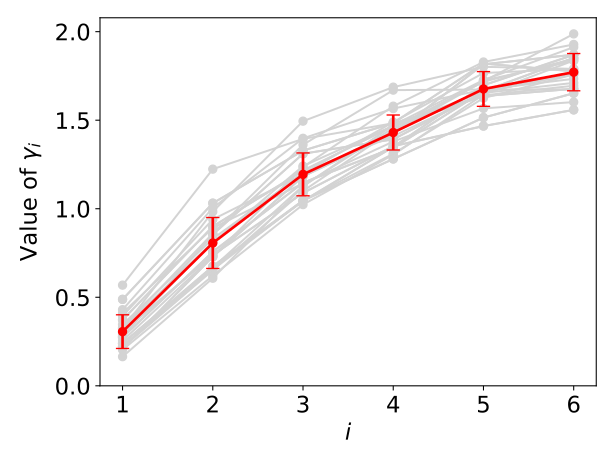} \\
        \ (a) & \ (b) \\
        \includegraphics[width=200pt]{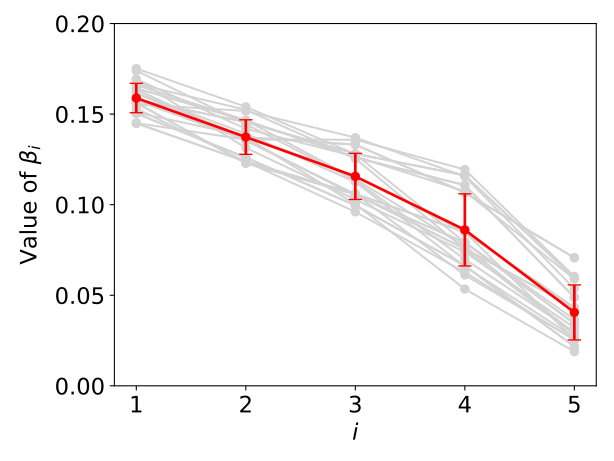} &
        \includegraphics[width=200pt]{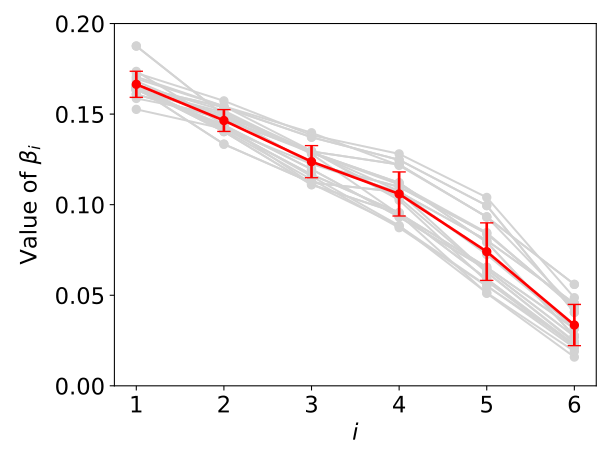} \\
        \ (c) & \ (d) \\
    \end{tabular}
    \caption{Correlation of optimal angles across rounds and instances. Plotted are the components of $\vec{\gamma}$ and $\vec{\beta}$ for optimal angles found on random graphs. In (a) and (b) we plot the components of $\vec{\gamma}$ for $p=5$ and $p=6$. Similarly in (c) and (d) we plot the components of $\vec{\beta}$ also for $p=5$ and $p=6$.}
    \label{fig:correlation}
\end{figure*}

\subsection{Interpolation-based Basin Hopping}
The plots in Figure~\ref{fig:examples} exhibit a pattern, which becomes more apparent with more vertices. Averaging the heat plots over many (i.e., 100) random graphs of 10 vertices, we can see the pattern much more clearly, as shown in Figure~\ref{fig:avg-heat}, for both the complete graph mixer and the ring mixer. These plots exhibit a large expectation value in the lower left hand corner, showing there exists a strong correlation between the optimal angles for different problem instances. 

Using basin-hopping optimization over the entire search space (i.e., not just for the first round), patterns in the optimal angles of $\vec{\gamma}$ and $\vec{\beta}$ emerged, all located in this lower left hand corner. The components of $\vec{\gamma}$ and $\vec{\beta}$ for different levels of $p$ are shown in Figure~\ref{fig:correlation}. In fact, we can clearly see a trend of how optimal angles evolve in the different rounds as they work their way to $p$. Optimum angles, thus,  are correlated across different stages (the $x$-axis in Figure~\ref{fig:correlation}) as well as across different graphs as illustrated by the different plot lines in the figure. 
Similar patterns in optimal angles have also been found in the problem of Max-Cut on regular graphs~\cite{zhou2018quantum, crooks2018performance}. 

We can see an analogy between these optimal angles and the quantum annealing algorithm (QAA), where the mixing Hamiltonian is slowly being turned off ($\beta$ angles are decreasing in magnitude), while the phase separating Hamiltonian is slowly being turned on ($\gamma$ angles are increasing in magnitude) as we transition from the ground state of the mixing Hamiltonian to the ground state of the phase separating Hamiltonian.

\begin{figure}
    \centering
    \includegraphics[scale=0.4]{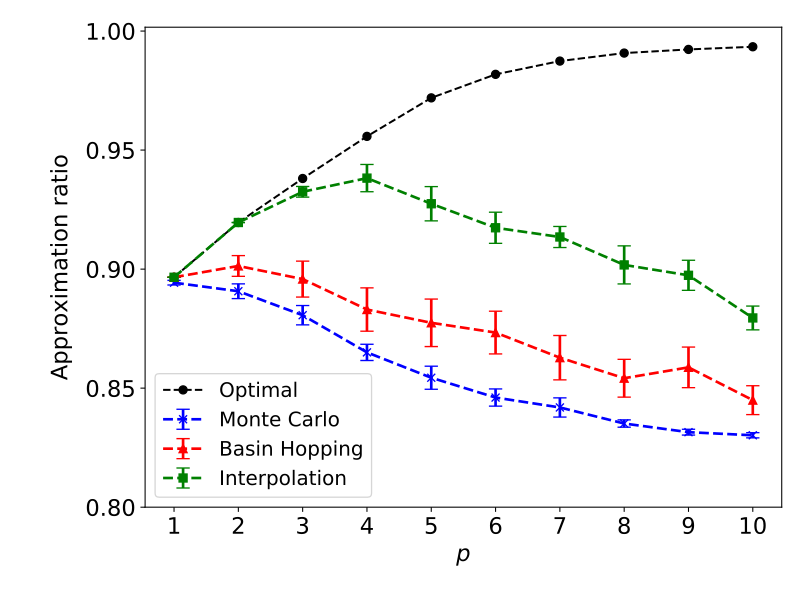}
    \caption{Comparison of angle selection strategies using a constant of number samples in each round, where the optimal approximation ratio is also plotted.}
    \label{fig:angle-selection}
\end{figure}

The pattern in Figure~\ref{fig:correlation} suggests an angle selection strategy which takes advantage of the overlap in optimal angles for different problem instances. To demonstrate this point, we compare the angle selection strategies of Monte Carlo sampling, basin-hopping optimization with random starting values, and basin-hopping optimization starting from a linear interpolation of the optimal angles in Figure~\ref{fig:correlation}. To make the comparison fair, all these strategies were restricted to the same number of samples of $F_p(\vec{\gamma}, \vec{\beta})$, and the number of samples was held constant for all $p$. The results of this comparison are in Figure~\ref{fig:angle-selection}, with a line showing the optimal approximation ratio. Clearly the `interpolation' strategy performs the best, reaching its peak performance at $p=4$. We again conjecture that the relative advantage of interpolation-based basin hopping will increase for larger graphs.

\section{Conclusion}
In conclusion, the Dicke states improve performance for both the complete graph and ring mixer, and in particular the Dicke states with the complete graph mixer performs better than the Dicke states with the ring mixer. For implementation on an all-to-all qubit device, the ring mixer can be implemented in depth $O(\log n)$ with no error, while the complete graph mixer can be implemented in query complexity $O(n^2/\sqrt{\delta})$ with error $\delta$. On more realistic architectures such as LNN connectivity, the ring mixer can be implemented exactly in $O(n)$ depth, the same depth as preparation of Dicke states.

Whether the trade-off between the improved performance but longer circuit depth of the complete graph mixer is worth the shorter depth but not-as-great performance of the ring mixer is still an open question. Using Monte Carlo samples, we saw the standard deviation of distribution of solutions decreases exponentially, making it exponentially harder to sample a better solution in deeper rounds of the algorithm. However, thankfully for the QAOA, optimal angles share patterns similar to the QAA, making it easier to find better solutions in deeper rounds of the algorithm.

Our results give insights into the structure of QAOA for a specific optimization problem, namely  Maximum $k$-Vertex Cover. While some of these insights have been found independently for the well-studied MaxCut QAOA (such as the correlation of angles and to some extent the mixer comparison), showing that similar results are valid across other optimization problems further confirms these insights. We have mostly refrained from explicitly expressing performance gains in quantitative terms as of course our insights into the algorithmic superiority of various QAOA components over others will have practical value only once we can treat larger graphs. Studying how the relative performance evolves for larger graphs is future work as is studying QAOA performance on other optimization problems.

\bibliographystyle{IEEEtran}
\bibliography{sources.bib}

\begin{thebibliography}{10}
\providecommand{\url}[1]{#1}
\csname url@samestyle\endcsname
\providecommand{\newblock}{\relax}
\providecommand{\bibinfo}[2]{#2}
\providecommand{\BIBentrySTDinterwordspacing}{\spaceskip=0pt\relax}
\providecommand{\BIBentryALTinterwordstretchfactor}{4}
\providecommand{\BIBentryALTinterwordspacing}{\spaceskip=\fontdimen2\font plus
\BIBentryALTinterwordstretchfactor\fontdimen3\font minus
  \fontdimen4\font\relax}
\providecommand{\BIBforeignlanguage}[2]{{%
\expandafter\ifx\csname l@#1\endcsname\relax
\typeout{** WARNING: IEEEtran.bst: No hyphenation pattern has been}%
\typeout{** loaded for the language `#1'. Using the pattern for}%
\typeout{** the default language instead.}%
\else
\language=\csname l@#1\endcsname
\fi
#2}}
\providecommand{\BIBdecl}{\relax}
\BIBdecl

\bibitem{farhi2014quantum}
E.~Farhi, J.~Goldstone, and S.~Gutmann, ``{A Quantum Approximate Optimization
  Algorithm},'' \emph{arXiv e-prints}, 2014,
  \href{https://arxiv.org/abs/1411.4028}{arXiv:1411.4028}.

\bibitem{farhi2014-e3lin2}
------, ``{A Quantum Approximate Optimization Algorithm Applied to a Bounded
  Occurrence Constraint Problem},'' \emph{arXiv e-prints}, 2014,
  \href{https://arxiv.org/abs/1412.6062}{arXiv:1412.6062}.

\bibitem{barak2015beating}
B.~Barak, A.~Moitra, R.~O’Donnell, P.~Raghavendra, O.~Regev, D.~Steurer,
  L.~Trevisan, A.~Vijayaraghavan, D.~Witmer, and J.~Wright, ``{Beating the
  Random Assignment on Constraint Satisfaction Problems of Bounded Degree},''
  in \emph{International Conference on Approximation Algorithms for
  Combinatorial Optimization Problems, APPROX'15}, ser. LIPICS, vol.~40, 2015,
  pp. 110--123,
  \href{https://doi.org/10.4230/LIPIcs.APPROX-RANDOM.2015.110}{doi:10.4230/LIPIcs.APPROX-RANDOM.2015.110}.

\bibitem{hadfield2019quantum}
S.~Hadfield, Z.~Wang, B.~O’Gorman, E.~G. Rieffel, D.~Venturelli, and
  R.~Biswas, ``{From the Quantum Approximate Optimization Algorithm to a
  Quantum Alternating Operator Ansatz},'' \emph{Algorithms}, vol.~12, no.~2,
  p.~34, 2019,
  \href{https://doi.org/doi:10.3390/a12020034}{doi:10.3390/a12020034}.

\bibitem{nasa}
Z.~Wang, N.~C. Rubin, J.~M. Dominy, and E.~G. Rieffel, ``{$XY$ mixers:
  Analytical and numerical results for the quantum alternating operator
  ansatz},'' \emph{Physical Review A}, vol. 101, no.~1, p. 012320, 2020,
  \href{https://doi.org/10.1103/PhysRevA.101.012320}{doi:10.1103/PhysRevA.101.012320}.

\bibitem{manurangsi2019}
P.~Manurangsi, ``{A Note on Max k-Vertex Cover: Faster FPT-AS, Smaller
  Approximate Kernel and Improved Approximation},'' in \emph{2nd Symposium on
  Simplicity in Algorithms, SOSA'19}, ser. {OASICS}, vol.~69, 2019, pp.
  15:1--15:21,
  \href{https://doi.org/10.4230/OASIcs.SOSA.2019.15}{doi:10.4230/OASIcs.SOSA.2019.15}.

\bibitem{petrank1994hardness}
E.~Petrank, ``{The hardness of approximation: Gap location},''
  \emph{Computational Complexity}, vol.~4, no.~2, pp. 133--157, 1994,
  \href{https://doi.org/10.1007/BF01202286}{doi:10.1007/BF01202286}.

\bibitem{ugc}
S.~Khot, ``{On the Power of Unique 2-Prover 1-Round Games},'' in \emph{34th
  Annual ACM Symposium on Theory of Computing, STOC'02}, 2002, pp. 767--775,
  \href{https://doi.org/10.1145/509907.510017}{doi:10.1145/509907.510017}.

\bibitem{saleem2020max}
Z.~H. Saleem, ``Max-independent set and the quantum alternating operator
  ansatz,'' \emph{International Journal of Quantum Information}, p. 2050011,
  2020,
  \href{https://doi.org/10.1142/S0219749920500112}{doi:10.1142/S0219749920500112}.

\bibitem{zhang2020qed}
Y.~Zhang, R.~Zhang, and A.~C. Potter, ``{QED driven QAOA for network-flow
  optimization},'' \emph{arXiv e-prints}, 2020,
  \href{https://arxiv.org/abs/2006.09418}{arXiv:2006.09418}.

\bibitem{fingerhuth2018quantum}
M.~Fingerhuth, T.~Babej \emph{et~al.}, ``{A quantum alternating operator ansatz
  with hard and soft constraints for lattice protein folding},'' \emph{arXiv
  e-prints}, 2018, \href{https://arxiv.org/abs/1810.13411}{arXiv:1810.13411}.

\bibitem{hodson2019portfolio}
M.~Hodson, B.~Ruck, H.~Ong, D.~Garvin, and S.~Dulman, ``{Portfolio rebalancing
  experiments using the Quantum Alternating Operator Ansatz},'' \emph{arXiv
  e-prints}, 2019, \href{https://arxiv.org/abs/1911.05296}{arXiv:1911.05296}.

\bibitem{arrasmith2020}
A.~Arrasmith, L.~Cincio, R.~D. Somma, and P.~J. Coles, ``{Operator Sampling for
  Shot-frugal Optimization in Variational Algorithms},'' \emph{arXiv e-prints},
  2020, \href{https://arxiv.org/abs/2004.06252}{arXiv:2004.06252}.

\bibitem{farhi2001quantum}
E.~Farhi, J.~Goldstone, S.~Gutmann, J.~Lapan, A.~Lundgren, and D.~Preda, ``A
  quantum adiabatic evolution algorithm applied to random instances of an
  np-complete problem,'' \emph{Science}, vol. 292, no. 5516, pp. 472--475,
  2001,
  \href{https://doi.org/10.1126/science.1057726}{doi:10.1126/science.1057726}.

\bibitem{bartschi2019dicke}
A.~B{\"a}rtschi and S.~Eidenbenz, ``{Deterministic Preparation of Dicke
  States},'' in \emph{22nd International Symposium on Fundamentals of
  Computation Theory, FCT'19}, 2019, pp. 126--139,
  \href{https://doi.org/10.1007/978-3-030-25027-0_9}{doi:10.1007/978-3-030-25027-0\_9}.

\bibitem{Mosca2001}
M.~Mosca and P.~Kaye, ``{Quantum Networks for Generating Arbitrary Quantum
  States},'' in \emph{Optical Fiber Communication Conference and International
  Conference on Quantum Information ICQI}, Jun 2001, p. PB28,
  \href{https://doi.org/10.1364/ICQI.2001.PB28}{doi:10.1364/ICQI.2001.PB28}.

\bibitem{Childs2002}
A.~M. Childs, E.~Farhi, J.~Goldstone, and S.~Gutmann, ``Finding cliques by
  quantum adiabatic evolution,'' \emph{Quantum Information \& Computation},
  vol.~2, no.~3, pp. 181--191, Apr 2002,
  \href{https://doi.org/10.26421/QIC2.3}{doi:10.26421/QIC2.3}.

\bibitem{wang2009}
H.~Wang, S.~Ashhab, and F.~Nori, ``{Efficient quantum algorithm for preparing
  molecular-system-like states on a quantum computer},'' \emph{Physical Review
  A}, vol.~79, no.~4, p. 042335, 2009,
  \href{https://doi.org/10.1103/PhysRevA.79.042335}{doi:10.1103/PhysRevA.79.042335}.

\bibitem{Wstates-logdepth}
D.~Cruz, R.~Fournier, F.~Gremion, A.~Jeannerot, K.~Komagata, T.~Tosic,
  J.~Thiesbrummel, C.~L. Chan, N.~Macris, M.-A. Dupertuis, and
  C.~Javerzac-Galy, ``{Efficient Quantum Algorithms for GHZ and W States, and
  Implementation on the IBM Quantum Computer},'' \emph{Advanced Quantum
  Technologies}, vol.~2, no. 5--6, p. 1900015, 2019,
  \href{https://doi.org/10.1002/qute.201900015}{doi:10.1002/qute.201900015}.

\bibitem{johnson-eigenvalues}
A.~E. Brouwer, S.~M. Cioab{\u{a}}, F.~Ihringer, and M.~McGinnis, ``{The
  smallest eigenvalues of Hamming graphs, Johnson graphs and other
  distance-regular graphs with classical parameters},'' \emph{Journal of
  Combinatorial Theory, Series B}, vol. 133, pp. 88--121, 2018,
  \href{https://doi.org/10.1016/j.jctb.2018.04.005}{doi:10.1016/j.jctb.2018.04.005}.

\bibitem{berry2009black}
D.~W. Berry and A.~M. Childs, ``{Black-Box Hamiltonian Simulation and Unitary
  Implementation},'' \emph{Quantum Information \& Computation}, vol.~12, no.
  1--2, pp. 29--62, 2012,
  \href{https://doi.org/10.26421/QIC12.1-2}{doi:10.26421/QIC12.1-2}.

\bibitem{wang2018fermionic}
Z.~Wang, S.~Hadfield, Z.~Jiang, and E.~G. Rieffel, ``{Quantum approximate
  optimization algorithm for MaxCut: A fermionic view},'' \emph{Physical Review
  A}, vol.~97, p. 022304, 2018,
  \href{https://doi.org/10.1103/PhysRevA.97.022304}{doi:10.1103/PhysRevA.97.022304}.

\bibitem{olson2012basin}
B.~Olson, I.~Hashmi, K.~Molloy, and A.~Shehu, ``{Basin Hopping as a General and
  Versatile Optimization Framework for the Characterization of Biological
  Macromolecules},'' \emph{Advances in Artificial Intelligence}, vol. 2012, p.
  674832, 2012,
  \href{https://doi.org/10.1155/2012/674832}{doi:10.1155/2012/674832}.

\bibitem{zhou2018quantum}
L.~Zhou, S.-T. Wang, S.~Choi, H.~Pichler, and M.~D. Lukin, ``{Quantum
  Approximate Optimization Algorithm: Performance, Mechanism, and
  Implementation on Near-Term Devices},'' \emph{Physical Review X}, vol.~10,
  no.~2, p. 021067, 2020,
  \href{https://doi.org/10.1103/PhysRevX.10.021067}{doi:10.1103/PhysRevX.10.021067}.

\bibitem{mbeng2019quantum}
G.~B. Mbeng, R.~Fazio, and G.~Santoro, ``{Quantum Annealing: a journey through
  Digitalization, Control, and hybrid Quantum Variational schemes},''
  \emph{arXiv e-prints}, 2019,
  \href{https://arxiv.org/abs/1906.08948}{arXiv:1906.08948}.

\bibitem{yang2017optimizing}
Z.-C. Yang, A.~Rahmani, A.~Shabani, H.~Neven, and C.~Chamon, ``{Optimizing
  Variational Quantum Algorithms Using Pontryagin’s Minimum Principle},''
  \emph{Physical Review X}, vol.~7, no.~2, p. 021027, 2017,
  \href{https://doi.org/10.1103/PhysRevX.7.021027}{doi:10.1103/physrevx.7.021027}.

\bibitem{brady2020optimal}
L.~T. Brady, C.~L. Baldwin, A.~Bapat, Y.~Kharkov, and A.~V. Gorshkov,
  ``{Optimal Protocols in Quantum Annealing and QAOA Problems},'' \emph{arXiv
  e-prints}, 2020, \href{https://arxiv.org/abs/2003.08952}{arXiv:2003.08952}.

\bibitem{biamonte2020}
V.~Akshay, H.~Philathong, M.~E.~S. Morales, and J.~D. Biamonte, ``{Reachability
  Deficits in Quantum Approximate Optimization},'' \emph{Physical Review
  Letters}, vol. 124, no.~9, p. 090504, 2020,
  \href{https://doi.org/10.1103/PhysRevLett.124.090504}{doi:10.1103/PhysRevLett.124.090504}.

\bibitem{baertschi2020}
A.~B{\"a}rtschi and S.~Eidenbenz, ``{Grover Mixers for QAOA: Shifting
  Complexity from Mixer Design to State Preparation},'' in \emph{IEEE
  International Conference on Quantum Computing \& Engineering, QCE'20}, 2020,
  \href{https://arxiv.org/abs/2006.00354}{arXiv:2006.00354}.

\bibitem{grover_fixed_point}
L.~K. Grover, ``{Fixed-Point Quantum Search},'' \emph{Physical Review Letters},
  vol.~95, no.~15, p. 150501, 2005,
  \href{https://doi.org/10.1103/PhysRevLett.95.150501}{doi:10.1103/PhysRevLett.95.150501}.

\bibitem{yoder2014fixed}
T.~J. Yoder, G.~H. Low, and I.~L. Chuang, ``{Fixed-Point Quantum Search with an
  Optimal Number of Queries},'' \emph{Physical Review Letters}, vol. 113,
  no.~21, p. 210501, 2014,
  \href{https://doi.org/10.1103/PhysRevLett.113.210501}{doi:10.1103/PhysRevLett.113.210501}.

\bibitem{crooks2018performance}
G.~E. Crooks, ``{Performance of the Quantum Approximate Optimization Algorithm
  on the Maximum Cut Problem},'' \emph{arXiv e-prints}, 2018,
  \href{https://arxiv.org/abs/1811.08419}{arXiv:1811.08419}.

\end{thebibliography}
\end{document}